\documentclass[a4paper,11pt]{article}
\usepackage{float}
\usepackage[pdftex]{graphicx}
\usepackage{subcaption}
\usepackage[T1]{fontenc}
\usepackage{lmodern}
\usepackage{slashed}
\usepackage[utf8]{inputenc}
\usepackage[english]{babel}
\usepackage{cite}
\usepackage{amsmath,amssymb,amsfonts,amsthm}
\usepackage{mathtools,mathrsfs,calligra,aurical}
\usepackage[nottoc,notlot,notlof]{tocbibind}
\usepackage{upgreek}
\usepackage{mathtools}
\numberwithin{equation}{section}
\allowdisplaybreaks
\usepackage[all]{xy}
\usepackage{xcolor}
\usepackage{graphicx}
\graphicspath{{images/}}
\usepackage{geometry}
\usepackage[toc,page]{appendix}
\usepackage{hyperref}
\usepackage[normalem]{ulem}
\hypersetup{colorlinks = true, linkcolor = red, linktocpage = true, citecolor = blue}

\usepackage{bm}
\usepackage{ragged2e}
\usepackage{appendix}
\usepackage{slashed}
\usepackage{bbold}
\usepackage{cancel}

\newcommand{\be}{\begin{equation}}
\newcommand{\bea}{\begin{eqnarray}}

\newcommand{\ee}{\end{equation}}
\newcommand{\eea}{\end{eqnarray}}

\DeclareMathAlphabet{\mathpzc}{OT1}{pzc}{m}{it}

\global\long\def\dd{\mathrm{d}}%
\usepackage{setspace}

\begin{document}
\begin{titlepage}
\begin{flushright}
\par\end{flushright}
\vskip 0.5cm
\begin{center}
\textbf{\LARGE \bf  Phase Transitions and Black Hole Stability in Gauged $\mathcal{N}=8$ Supergravity}\\
\vskip 5mm

\vskip 1cm

\large {\bf Andr\'{e}s Anabal\'{o}n}$^{~a ~b ~c}$\footnote{anabalo@gmail.com}, \large {\bf Dumitru Astefanesei}$^{~d}$\footnote{dastefanesei@gmail.com}, \large {\bf Julio Oliva}$^{~a}$\footnote{juoliva@udec.cl}, \large {\bf Gabriel Ortega}$^{~a}$\footnote{gortega2020@udec.cl} and \large {\bf Jorge Urbina}$^{~a}$\footnote{jurbina2020@udec.cl}

\vskip .5cm 

$^{(a)}${\textit{Departamento de Física, Universidad de Concepción, Casilla 160-C, Concepción, Chile.}}\\ \vskip .1cm
$^{(b)}${\textit{Instituto de F\'isica Te\'orica, UNESP-Universidade Estadual Paulista \\
R. Dr. Bento T. Ferraz 271, Bl. II, Sao Paulo 01140-070, SP, Brazil.}}\\ \vskip .1cm
$^{(c)}${\textit{Max Planck Institute for Gravitational Physics, Am Mühlenberg 1, D-14476 Potsdam, Germany}}\\ \vskip .1cm
$^{(d)}${\textit{Pontificia Universidad Católica de Valparaíso, Instituto de Física, Av. Brasil 2950, Valparaíso, Chile.}}
\end{center}
\begin{abstract}
{It is well known that there is a region of parameter space where all purely electric, static, dilatonic black holes are unstable within the STU models of maximal supergravity. We show that, for planar black holes, it is possible to complete the thermal phase space with AdS solitons, in such a way that the instability of the black holes signals the onset of confinement in the dual field theory. The analysis is done for the $D=4$ STU model of maximal gauged supergravity which naturally uplift to M-theory on the $S^7$. }
\end{abstract}

\vfill{}
\vspace{1.5cm}
\end{titlepage}

\setcounter{footnote}{0}
\tableofcontents

\section{Introduction and Discussion}

Black holes are very interesting objects. Despite being macroscopical, the no-hair theorems make them kind of fundamental as they can be described with a small set of parameters. Furthermore, via holography they might describe thermal states of a large $N$, strongly coupled, quantum field theory. A necessary condition for thermal states to belong to a thermodynamic ensemble is that they be locally stable; otherwise, they are not thermal states in equilibrium. It is well-known that there is a region of the parameter space where static dilatonic black holes are never locally stable within the STU models of the maximal gauged supergravities \cite{Duff:1999gh, Cvetic:1999ne, Cvetic:1999rb, Gubser:2000ec, Gubser:2000mm}. To achieve a complete understanding of the fate of these thermal states through holography remains an important open problem, for recent work see \cite{Henriksson:2019zph, Gladden:2024ssb, OyarzoCatalan:2025fah, Anabalon:2024cnb, Anabalon:2025uzl,Buchel:2025jup, Buchel:2025ves,Gladden:2025glw,Anabalon:2024lgp,Dias:2024edd}.

Recently, a new family of supersymmetric solitons has been shown to exist within different truncations of the STU models of maximal gauged supergravities
\cite{Anabalon:2021tua, Anabalon:2022aig, Anabalon:2023oge,Anabalon:2024qhf,Anabalon:2024che}. These are generalizations of the AdS soliton geometry found in \cite{Horowitz:1998ha}. These are horizonless geometries that are well suited to describe holographic confinement, for recent works see \cite{Nunez:2023xgl,Fatemiabhari:2024aua,Chatzis:2024top,Chatzis:2024kdu,Giliberti:2024eii,Castellani:2024ial,Fatemiabhari:2024lct,Jokela:2025cyz,Nunez:2025gxq,Chatzis:2025dnu,Macpherson:2025pqi,Nunez:2025ppd,Chatzis:2025hek,Fatemiabhari:2025usn}. 

Indeed, the fact that the Hawking-Page phase transition does not exist for planar black holes in AdS, make the AdS soliton geometries the natural confining geometry in this case \cite{Surya:2001vj,Nishioka:2009zj,Horowitz:2010jq,Anabalon:2022ksf,Quijada:2023fkc}. Hence, it is of interest to investigate how these geometries fit into the phenomenology of maximal supergravities. In particular, in this article we show that they provide good candidates for the end state of the instability of the electrically charged black holes we just mentioned.

Thus, we focus on a purely dilatonic truncation of gauged $\mathcal{N}=8$ supergravity, where the axions are set to zero. Specifically, we explore the so-called $\text{T}^3$ model of gauged $\mathcal{N}=8$ supergravity model in four dimensions. This truncation effectively reduces the dynamics to a single ``effective'' scalar field coupled to two independent charges. In the grand canonical ensemble, where the temperature and the chemical potentials associated with the gauge fields are held fixed at the boundary, we demonstrate that, for a given set of boundary data, there are several possible black hole solutions. This analysis allows us to identify the spinodal curves in the parameter space, which delimits the region where the black hole solutions become unstable. We show that it is possible to provide a simple and rather straightforward parameterization of the phase space that yield stable black hole solutions. The regions of local stability are plotted and explicitly found in the canonical and grand canonical ensembles.

Moreover, we investigate the global stability of the system by allowing phase transitions towards a different background configuration, the planar hairy soliton. The solitons of this model were constructed in detail in \cite{Anabalon:2022aig}. We show that there are first order phase transition between the black hole and the soliton along the boundary of the region of stability. 

The outline of the article is as follows. In Section 2 we introduce the dilatonic sector of the STU model of $\mathrm{SO}(8)$ gauged $\mathcal{N}=8$ supergravity and specify the consistent truncation that allows us to focus on purely electric or magnetic configurations. In Section 3 we review the planar electrically charged dilatonic black hole solutions, analyze their asymptotic structure, identify the dual one-point functions, and derive the equation of state. We then study local thermodynamic stability in the microcanonical ensemble through the Hessian of the energy and determine the corresponding spinodal region.

In Section 4 we construct the grand canonical ensemble, express the physical parameters in terms of temperature and chemical potentials, and show that the solution space is governed by a cubic equation. We provide an explicit parametric description of the region of local stability in this ensemble. In Section 5 we analyze the full space of solutions, first discussing the thermodynamically stable black holes and their Gibbs free energy, and then describing the AdS soliton solutions with the same boundary conditions, including a simple parametric solution and their energy spectrum. Finally, in Section 6 we study the first-order phase transitions between black holes and solitons, determine the conditions under which each phase dominates, and describe the smooth continuation between both branches at the boundary of the black hole stability region.

\section{The model}\label{sec:model}
We shall consider the dilatonic sector of the STU\ model of the
$\mathrm{SO}(8)$-gauged, $\mathcal{N}=8$ supergravity with bulk action:
\begin{equation}
\mathcal{S}=\frac{1}{2\,\kappa}\int\! d^{4}x\:\sqrt{-g}\left(R+\sum_{a=1}^{3}- \frac{\left(\partial\Phi_{a}\right)  ^{2}}{2}+\frac{2}{L^{2}}\,\cosh\left(  \Phi_{a}\right)
-\frac{1}{4}\:\sum_{\Lambda=1}^{4}X_{\Lambda}^{-2}F_{\Lambda}^{2}\right)\:,
\label{LSTU}
\end{equation}
where 
\begin{align}
F_{\Lambda}=\dd A_{\Lambda}\,,
\qquad
X_{\Lambda}=e^{-\frac{1}{2}\vec{a}_{\Lambda}\cdot
\vec{\Phi}}\,,
\qquad
\vec{\Phi}=\left(\Phi_{1},\Phi_{2},\Phi_{3}\right)\,,
\end{align}
and
\begin{equation}
\vec{a}_{1}=\left(1,1,1\right),
\qquad
\vec{a}_{2}=\left(1,-1,-1\right),
\qquad
\vec{a}_{3}=\left(-1,1,-1\right),
\qquad
\vec{a}_{4}=\left(-1,-1,1\right).\;
\end{equation}
We will consider either purely magnetic or purely electric solutions so is
consistent to truncate the axions to zero. This Lagrangian \eqref{LSTU} can be
obtained from the compactification of eleven dimensional supergravity over the
seven-sphere. An explicit ansatz for this oxidation can be found in \cite{Cvetic:1999xp}. 

\section{Electric dilatonic black holes}

In the case of a spherical horizon, these solutions were found in \cite{Duff:1999gh},
and their generalizations to planar and hyperbolic horizons was constructed
in \cite{Cvetic:1999xp}. Here, we present the planar solutions. With the coordinates of \cite{Anabalon:2024cnb}, the metric reads
\begin{align}
\dd s^{2} & =-\frac{f(r)}{\sqrt{H(r)}}\dd t^{2}+\frac{\sqrt{H(r)}}{f(r)}\dd r^{2}+\frac{r^{2}}{L^2}\sqrt{H(r)}\left(\dd x^{2}+\dd \varphi^{2}\right)\,,\\
f(r) & =\frac{r^{2}}{L^{2}}H(r)-\frac{m}{r}-\frac{q}{r^{2}}\,,\qquad H(r)=H_{1}H_{2}H_{3}H_{4}\,,\qquad H_{\Lambda}=1+\frac{q_{\Lambda}}{r}\,.\label{general electric background}
\end{align}
The dilatons and the gauge fields are
\begin{align}
\Phi_{1} & =\frac{1}{2}\log\left(\frac{H_{1}H_{2}}{H_{3}H_{4}}\right)\,,\qquad\Phi_{2}=\frac{1}{2}\log\left(\frac{H_{1}H_{3}}{H_{2}H_{4}}\right)\,,\qquad\Phi_{3}=\frac{1}{2}\log\left(\frac{H_{1}H_{4}}{H_{2}H_{3}}\right)\,,\\
A^{\Lambda} & =\left( \frac{Q_{\Lambda}}{rH_{\Lambda}} -\mu_{\Lambda}\right)\dd t\,,\qquad Q_{\Lambda}^{2}=q_{\Lambda}m-q\,.
\end{align}
Where $\mu_\Lambda$ yields the chemical potentials when it is fixed in such a way that the gauge fields are regular in the Euclidean continuation. 

\subsection{Asymptotic analysis}
Here all the indexes labeled with $\Lambda$ or $\Lambda_i$ run from $1$ to $4$. It is possible to check that the following asymptotic change of coordinates
\begin{align}
r=&\rho - \frac{1}{4}\sum_{\Lambda} q_\Lambda + \frac{1}{\rho}\left(\frac{3}{32}\sum_{\Lambda} q_\Lambda^2 - \frac{1}{16}\sum_{\Lambda_1<\Lambda_2} q_{\Lambda_1} q_{\Lambda_2}\right)\nonumber \\
&- \frac{1}{32\rho^2}(q_1 + q_2 - q_3 - q_4)(q_1 - q_2 + q_3 - q_4)(q_1 - q_2 - q_3 + q_4) +O(\rho^{-3})
\end{align}
yields the fall-off for the metric functions
\begin{align}
g_{x x} =&\,g_{\varphi \varphi} = \frac{r^2}{L^2}\sqrt{H(r)}
=\frac{\rho^2}{L^2}+O(\rho^{-2})\, , \\
-g_{t t} =& \frac{f(r)}{\sqrt{H(r)}} =\frac{\rho^2}{L^2}-\frac{m}{\rho}+O(\rho^{-2})\, , \\
g_{\rho \rho}  =& \frac{\sqrt{H(r)}}{f(r)}\left(\frac{\dd r}{\dd \rho}\right)^2 = \frac{L^2}{\rho^2} - \frac{L^2}{16\rho^4} \left( 3 \sum_{\Lambda} q_{\Lambda}^2 - 2 \sum_{\Lambda_1 < \Lambda_2} q_{\Lambda_1} q_{\Lambda_2} \right)\, \nonumber\\ 
&+ \frac{L^2}{8\rho^5} \left( \sum_{\Lambda} q_{\Lambda}^3 - \sum_{\Lambda_1 \neq \Lambda_2} q_{\Lambda_1}^2 q_{\Lambda_2} + 2 \sum_{\Lambda_1 < \Lambda_2 < \Lambda_3} q_{\Lambda_1} q_{\Lambda_2} q_{\Lambda_3} + 8mL^2 \right)+O(\rho^{-6})\, ,
\end{align}
and for the scalars we get
\begin{align}
\Phi_1 =& \frac{1}{2\rho}(q_1 + q_2 - q_3 - q_4) - \frac{1}{8\rho^2}(q_1 - q_2 - q_3 + q_4)(q_1 - q_2 + q_3 - q_4) + O(\rho^{-3})\, ,\\
\Phi_2 =&\frac{1}{2\rho}(q_1 - q_2 + q_3 - q_4) - \frac{1}{8\rho^2}(q_1 + q_2 - q_3 - q_4)(q_1 - q_2 - q_3 + q_4) + O(\rho^{-3})\, ,\\
\Phi_3 =&\frac{1}{2\rho}(q_1 - q_2 - q_3 + q_4) - \frac{1}{8\rho^2}(q_1 + q_2 - q_3 - q_4)(q_1 - q_2 + q_3 - q_4) + O(\rho^{-3})\, .
\end{align}
Therefore we see that the scalars yield the following VEVs
\begin{align}
z_1 = \frac{1}{2}(q_1 + q_2 - q_3 - q_4)\, ,\\
z_2 = \frac{1}{2}(q_1 - q_2 + q_3 - q_4)\, ,\\
z_3 = \frac{1}{2}(q_1 - q_2 - q_3 + q_4)\, ,
\end{align}
and the sub-leading terms of each of scalar, $w_i$ can be obtained as follows
\begin{equation}
w_i = -\frac{\partial W }{\partial z_ i}=-\frac{1}{2}\frac{\partial (z_1 z_2 z_3) }{\partial z_ i}
\end{equation}
Hence, these solutions have conformally invariant boundary conditions relevant for the description of the ABJM theory \cite{Freedman:2016yue}. 

For these boundary conditions the dual energy momentum tensor is given by \cite{Anabalon:2014fla}
\begin{equation}\label{EMtensor}
\left\langle T_{\varphi\varphi}\right\rangle =\frac{m}{2\,\kappa\, L^{2}}\,,
\qquad
\left\langle T_{x x}\right\rangle =\frac{m}{2\,\kappa\,L^{2}}\,,
\qquad
\left\langle T_{t t}\right\rangle =\frac{m}{\kappa\,L^{2}}\,,
\end{equation}
where we pick the following representative of the conformal boundary metric
\begin{equation}
\dd s_{\partial }^{2}=-\dd t^2+\dd x^2+\dd \varphi^2\, .
\end{equation}

\subsection{The equation of state and the Hessian}
\begin{figure}[h!]
    \centering
    \includegraphics[width=0.7\textwidth]{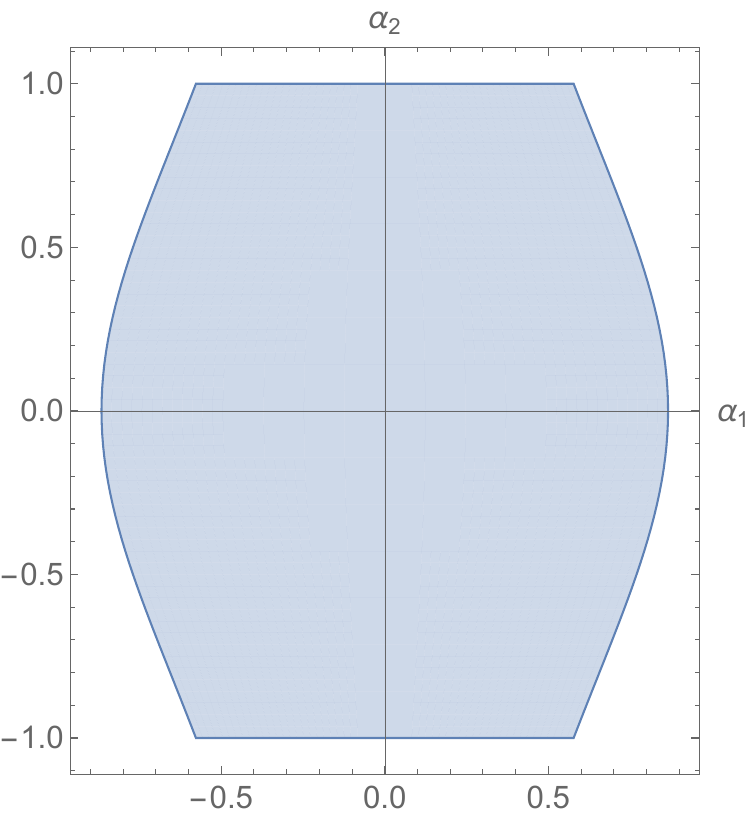}
    \caption{Stability region of the T${}^{3}$ model in the microcanonical ensemble.}
    \label{fig:stability}
\end{figure}

The equation of state of these solutions is given by the relation between the energy, the charges and the entropy. The entropy is proportional to the area of the horizon, which is proportional to the square root of the determinant of the constant $t-r$ sector of the metric evaluated at the location of the horizon. Hence, up to a numerical factor, the entropy is proportional to $A$
\begin{equation} \label{Area}
A=\frac{r_0^2}{L^2} \sqrt{H(r_0)
}\, ,
\end{equation}
where $f(r_0)=0$. Hence,
\begin{equation}
f(r_0)=\frac{L^2}{r_0^2}A^2-\frac{m}{r_0}-\frac{q}{r_0^2}=0\implies r_0=\frac{A^2L^2-q}{m}\, .
\end{equation}
This can be used to find that
\begin{equation}
H(r_0)=\frac{1}{(A^2L^2-q)^4} \prod_{\Lambda}\left(A^2L^2+Q_\Lambda^2\right)=\frac{1}{m^4r_0^4} \prod_{\Lambda}\left(A^2L^2+Q_\Lambda^2\right)\, .
\end{equation}
When this is replaced in \eqref{Area} we obtain the equation of state
\begin{equation}
m=\frac{1}{A^{1/2}L}\prod_{\Lambda}\left(A^2L^2+Q_\Lambda^2\right)^{1/4}\, .
\end{equation}
$m$ is proportional to the energy of the system. Hence, the condition of stability is that $m$ is a convex function of the variables $Y^{\lambda}=(AL,Q_{\Lambda})$\footnote{The thermodynamical variables are normalized such that they all have the same dimensions.}. The Hessian,
\begin{equation}
H_{\lambda \sigma}=\frac{\partial^2 m }{\partial Y^{\lambda} \partial Y^{\sigma}}\, ,
\end{equation}
is a $5\times 5$ matrix and the convexity of $m$ is ensured provided all the eigenvalues are non-negative. When $Q_4=Q_3=Q_2$, the determinant is
\begin{equation}\label{dethess}
\det H = \frac{(A^2 L^2 - Q_2^2)^2 \left(3 A^4 L^4 - 2 A^2 L^2 Q_1^2 - Q_1^2 Q_2^2\right)}%
{64\, A^{9/2} L^7\, (A^2 L^2 + Q_1^2)^{3/4}\, (A^2 L^2 + Q_2^2)^{9/4}}
\end{equation}
We find that the Hessian has a repeated eigenvalue, $\chi$, which constrains a charge
\begin{equation}\label{Q2}
\chi = \frac{(A^{2} L^{2}-Q_2^2) (A^{2} L^{2} + Q_{1}^{2})^{1/4}}
{2 L\, \sqrt{A}\, (A^{2} L^{2} + Q_{2}^{2})^{5/4}}>0\implies \alpha_2^2\equiv \frac{Q_2^2}{A^2L^2}<1\, .
\end{equation}
The other three eigenvalues are solutions of a cubic equation. A necessary condition for the non-negativity of all eigenvalues is that the determinant \eqref{dethess} is non-negative. This condition can be read as 
\begin{equation}\label{pos}
\alpha_1^2\equiv \frac{Q_1^2}{A^2L^2}<\frac{3}{2+\alpha_2^2}
\end{equation}
We checked that the cubic equation has only real positive roots provided \eqref{Q2} and \eqref{pos} are satisfied. We conclude that the stability region is the convex set bounded by the spinodal line of the figure \eqref{fig:stability}.

The variational problem in AdS fixes the conformal boundary metric at spacelike infinity. In the Euclidean formulation, this corresponds to fixing the temperature of the system. We also fix the boundary conditions for the gauge fields, which correspond to fix the chemical potential. Studying the phase space of the theory with these boundary conditions corresponds to work in the grand canonical ensemble. We now proceed to construct the solution space in this ensemble.

\section{Grand canonical ensemble}
To construct the grand canonical ensemble, we need to express the mass as a function of the temperature and the chemical potentials $\mu_{\Lambda}$. To do this, we introduce intermediate variables $(V, Z_{\Lambda})$ which, after imposing the relevant relations among the parameters of the system, become functions of the temperature and the chemical potentials. We find that, up to overall factors, the physical variables depend on the following ratios
\begin{equation}
\sigma_{\Lambda}=\frac{\mu_{\Lambda}}{2\pi T L }\, .
\end{equation}
We eliminate everywhere the harmonic functions evaluated at the horizon in terms of the other variables using the definition of the chemical potentials,
\begin{equation}
H_{\Lambda}(r_0)=\frac{Q_{\Lambda}}{r_0\mu_{\Lambda}}\, .    
\end{equation}
We postulate the following relation between the charges and the chemical potentials 
\begin{equation}
Q_{\Lambda}= 2\pi T V L^2  \mu_{\Lambda} Z_{\Lambda} \, ,
\end{equation}
with $q=(2\pi T)^4 V^4L^6-mr_0$. This simplifies the constraint $f(r_0)=0$ to $Z_1Z_2^3=1$. From the definition of the temperature, $T$, we can find the mass parameter
\begin{equation}
T=\frac{f'(r_0)}{4\pi H(r_0)^{1/2}}\implies m =L^4 (2\pi T)^3\left(Z_{2}^{2} (Z_{2} + 3 Z_{1}) V - 2\right)V^{2}
\end{equation}
The equations $Q_{\Lambda}^2-q_{\Lambda}m+q=0$ are
\begin{align}\label{EQ1}
3 V^{2} Z_{1}^{2} Z_{2}^{2} - 2 V Z_{1} - \sigma_{1}^{2} Z_{1}^{2}&=0\, , \\\label{EQ2}
V^{2} Z_{2}^{4} + 2 V^{2} - 2 V Z_{2} - \sigma_{2}^{2} Z_{2}^{2}&=0\, .
\end{align}
Equations \eqref{EQ1} and \eqref{EQ2} can be combined to yield a linear equation for the variable $V$,
\begin{equation} \label{V}
V=
-\frac{1}{2}\,
 \frac{\sigma_{1}^{2} Z_{2}^{4} + 2 \sigma_{1}^{2} - 3 Z_{2}^{4} \sigma_{2}^{2}}
      {Z_{2}^{3}(Z_{2}^{4}-1)}\, .
\end{equation}
Given \eqref{V}, \eqref{EQ1} and \eqref{EQ2} imply that
\begin{equation} \label{cubic}
4 Z_2^{12} \sigma_2^2 
+ \left( 6 \sigma_1^2 \sigma_2^2 - \sigma_1^4 - 9 \sigma_2^4 - 4 \sigma_1^2 - 4 \sigma_2^2 \right) Z_2^{8}
- 4 \sigma_1^2 \left( -3 \sigma_2^2 + \sigma_1^2 - 1 \right) Z_2^{4}
- 4 \sigma_1^{4}=0 \, .
\end{equation}
The cubic equation for $Z_2^4$ \eqref{cubic}, indicates that there is at most $3$ solutions for every value of the boundary conditions parameterized by $(\sigma_1, \sigma_2)$. 
\subsection{Stability Again}
To find the stability region for the variables $(\sigma_1, \sigma_2)$ we notice that 
\begin{equation}
\alpha_2=\frac{Q_2}{AL}=\sigma_2\frac{Z_2}{V}\, .
\end{equation}
Hence, we introduce the change of variables
\begin{equation}\label{new2}
\sigma_2=\nu_2 \frac{V}{Z_2}\, , 
\end{equation}
with $\nu_2 \in (-1,1)$ to ensure the condition of stability \eqref{Q2}. The other charge has a similar parametrization
\begin{equation}
\alpha_1=\frac{Q_1}{AL}=\frac{\sigma_1}{V Z_2^3}\, .
\end{equation}
The following change of parameters
\begin{equation}\label{new1}
\sigma_1=\frac{\sqrt{3}}{\sqrt{2+\nu^2_2}}\, \nu_1 V Z_2^{3}\, ,
\end{equation}
enforces the second condition of stability, \eqref{pos}, provided $\nu_1 \in (-1,1)$. If we replace \eqref{new2} and \eqref{new1} in \eqref{EQ1} and \eqref{EQ2} we get

\begin{align} \label{Z2}
Z_2^4 &= \frac{\nu_2^4+2+3\nu_2^2}{2+3\nu_1^2+\nu_2^2}\, , \\ \label{V0}
V&=-\frac{2}{3}\frac{Z_2(2+\nu_2^2)}{\nu_1^2 Z_2^4-2-\nu_2^2}\, .
\end{align}
We use \eqref{Z2} and \eqref{V0} to have the explicit form of $(\sigma_1,\sigma_2) $ in terms of the quantities bounded by the stability conditions, $\nu_{\Lambda}$. The result is
\begin{align}
\sigma_1&=-\frac{2}{3}\frac{\sqrt{3}\sqrt{2+\nu_2^2}\,\nu_1\,(\nu_2^2+1)}{-\nu_2^2+\nu_2^2\nu_1^2-2-2\nu_1^2}
\, , \\ 
\sigma_2&=-\frac{2}{3} \frac{(2 + 3 \nu_1^2 + \nu_2^2) \, \nu_2}{- \nu_2^2 + \nu_2^2 \nu_1^2 - 2 - 2 \nu_1^2}
\end{align}
\begin{figure}[h!]
    \centering
    \includegraphics[width=0.7\textwidth]{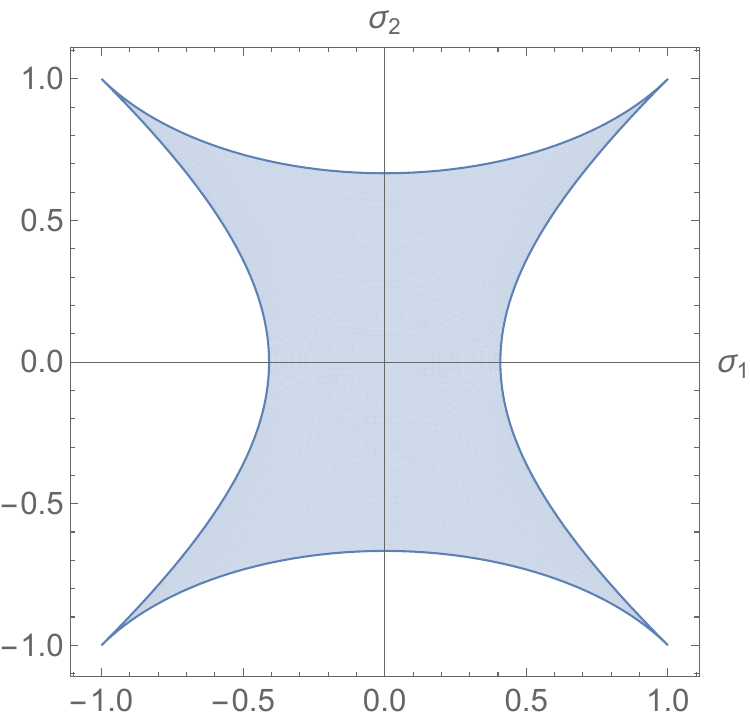}
    \caption{Boundary conditions for AdS solitons in the T${}^{3}$ model.}
    \label{fig:stabilitygc}
\end{figure}
The boundary conditions that yield stable black holes correspond to the shaded region shown in the figure \eqref{fig:stabilitygc}. We note that this analysis yield a relatively simple parametric solution to the cubic equation \eqref{cubic}.

\section{The space of solutions}
\subsection{Black Holes}
\begin{figure}[h!]
    \centering
    \includegraphics[width=0.7\textwidth]{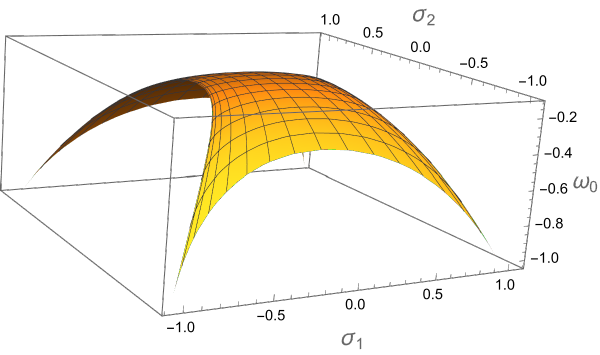}
    \caption{Free energy density of the stable solutions of the T${}^{3}$ model in the grand canonical ensemble. Stable thermal states exist for $\omega_0\in [-1,-\frac{4}{27}]$.}
    \label{fig:free}
\end{figure}

The thermodynamically stable solutions are characterized by their free energy. Hence, we plot the Gibbs free energy density $\frac{L^2}{\kappa} \omega$ versus $\sigma_{\Lambda}$. The standard thermodynamical relations allow an straightforward calculation. Indeed,
\begin{equation}
\omega = -\frac{\kappa}{L^2} P =-\frac{m}{2L^4}=-\frac{1}{2} (2\pi T)^3\left(Z_{2}^{2} (Z_{2} + 3 Z_{1}) V - 2\right)V^{2}\, .
\end{equation}
where $P$ is the pressure of the dual fluid given in \eqref{EMtensor}. The dimensionless free energy $\omega_0=\frac{\omega}{(2\pi T)^3}$ vs $(\sigma_1, \sigma_2)$ can be seen in figure \eqref{fig:free}. We learn that stable thermal states exist for $\omega_0\in [-1,-\frac{4}{27}]$.

\subsection{AdS Solitons}
\begin{figure}[h!]
    \centering
    \includegraphics[width=0.7\textwidth]{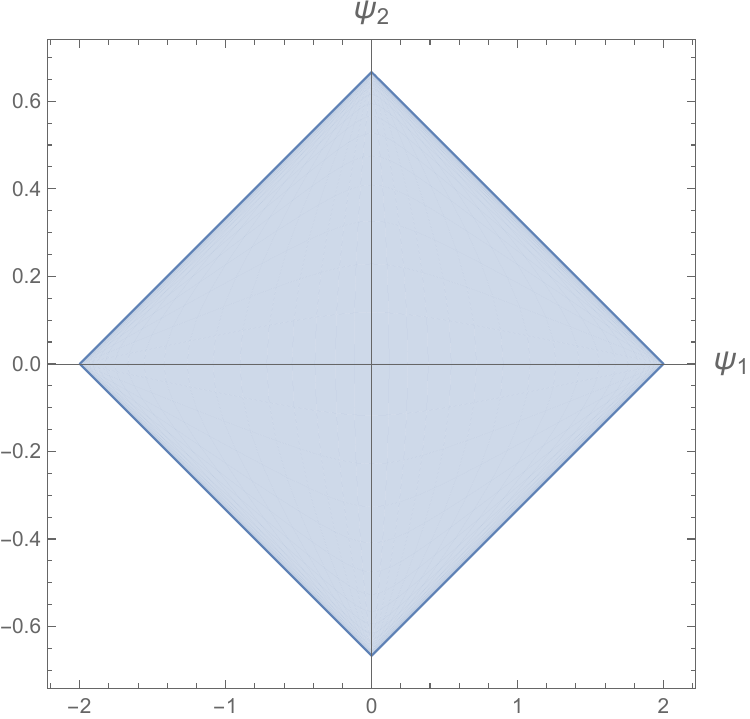}
    \caption{The boundary conditions that yield solitons for the T$^3$ model.}
    \label{fig:solitons1}
\end{figure}

If the metric of the conformal boundary has a compact direction, $\varphi \in [0,\Delta]$, there exist solitons that have the same boundary conditions of the black holes. These solitons are in equilibrium with an arbitrary chemical potentials because loops around the time direction are non-contractible, and therefore time is a globally defined coordinate. These solutions are described in detail in \cite{Anabalon:2022aig}. The solution space is characterized by the Wilson loops
\begin{equation}
\psi_1= \frac{1}{2\pi L}\,\lim_{r \rightarrow \infty}\oint A^{1}_{\varphi}\dd \varphi \, ,
\qquad
\psi_2=\frac{1}{2\pi L}\,\lim_{r \rightarrow \infty}\oint A^{2}_{\varphi} \dd \varphi \, , 
\end{equation}
with the energy density $\frac{L^2}{\kappa}\rho_{sol}$ \footnote{Note that $\psi^{\rm Here}_1=\sqrt{2}\,\psi^{\rm There}_1$ and $\psi^{\rm Here}_2=\sqrt{\frac{2}{3}}\,\psi^{\rm There}_2$ where ``There'' are the variables of \cite{Anabalon:2022aig}.}
\begin{equation}
\rho_{sol}= \pm 
\frac{2\pi^{3}}{\Delta^{3}}\,
x_0 \,\left| 2 x_{0}^{2} \psi_{1}^{2} + \psi_{1}^{2} - 3 \psi_{2}^{2} \right|\,
\frac{ \psi_{1}^{2} x_{0}^{4} -  \psi_{2}^{2} }{ (x_{0}^2-1)^{2}  }\,
\end{equation}
where the $+$ sign is for solutions with $x_0>1$ and the $-$ sign is for solutions with $x_0<1$. The solutions are the roots of the polynomial 
\begin{equation}\label{cubicsol}
4\psi_{1}^{4} x_{0}^{6}
+ 4\psi_{1}^{2}(\psi_{1}^{2} - 3\psi_{2}^{2} + 1)x_{0}^{4}
+ (\psi_{1}^{4} - 6\psi_{1}^{2}\psi_{2}^{2} - 4\psi_{1}^{2} - 4\psi_{2}^{2} + 9\psi_{2}^{4})x_{0}^{2}
+ 4\psi_{2}^{2}=0
\end{equation}
Inspired by the parametric solution of the equation for the existence of black holes \eqref{cubic}, we found the following simple parametric solution of the equation \eqref{cubicsol},
\begin{align}\label{solsimple}
x_0 &= \frac{\cosh{\xi_2}}{\cosh{\xi_1}}\, ,\\ \label{psi1}
\psi_{1}&=\frac{\sinh{2\xi_1}}{\cosh{\xi_1}^2+3\cosh{\xi_2}^2-1}
\, ,\\\label{psi2}
\psi_{2}
&=\frac{\sinh{2\xi_2}}{\cosh{\xi_1}^2+3\cosh{\xi_2}^2-1}\, .
\end{align}
When \eqref{psi1} and \eqref{psi2} are replaced in the equation \eqref{cubicsol}, it factorizes. In addition to the solution \eqref{solsimple} we find that there is always another solution with $x_0^2>0$. Therefore, there are two soliton solutions at each value of the boundary conditions $(\psi_1,\psi_2)$. The set of boundary conditions that yield these solutions can be seen in figure \eqref{fig:solitons1}. There is also a branch of solitons with positive energy that is not relevant for the possible phase transitions with the black holes.

\begin{figure}[h!]
    \centering
    \includegraphics[width=0.7\textwidth]{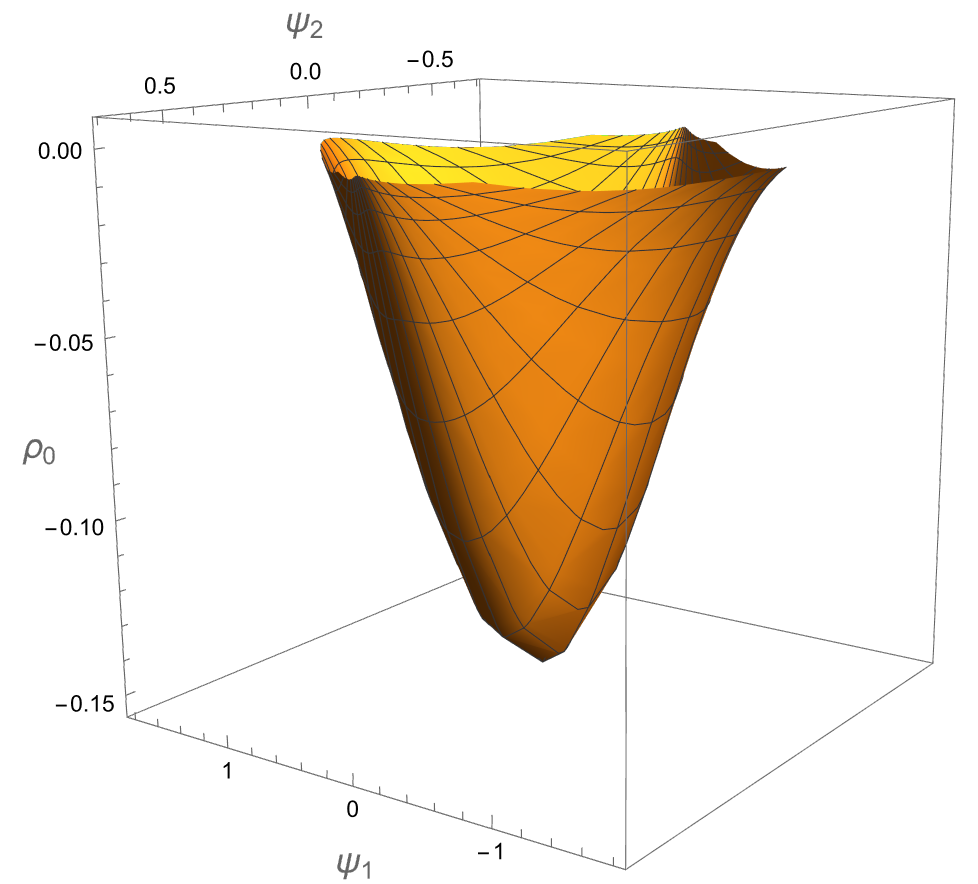}
    \caption{Energy density of the branch of solitons with negative energy. We plot $\rho_0=\frac{\Delta^3}{(2\pi)^3}\rho_{sol}\in [-\frac{4}{27},0]$. The solutions at $\rho_0=0$ are supersymmetric.}
    \label{fig:energysoliton}
\end{figure}
\section{First order phase transitions}
There is a phase transition between the soliton and the black hole if 
\begin{equation}
\omega =\rho_{sol}\implies T^3 \Delta^3 =\frac{\rho_0}{\omega_0}\, .
\end{equation}
The soliton is thermodynamically favorable if 
\begin{equation}
T^3 \Delta^3 >\frac{\rho_0}{\omega_0} \, .
\end{equation}
Hence, hot black holes can decay into a small soliton and cold black holes can decay into a large or a small soliton provided this inequality is satisfied. This is a first order phase transition as the entropy goes from being very large in the black hole to $0$ in the soliton.

At the boundary of the region of stability of the black holes one can then smoothly continue the space of solutions from black holes to solitons. There are two cases. 

$\mathbf{\nu_2^2=1}$. In this case
\begin{equation}
\omega_0(\nu^2_2=1)=-\frac{8}{27}\,\frac{\sqrt{6}\,\left(3+3\nu_1^2\right)^{5/2}}{\left(\nu_1^2+3\right)^3}
\end{equation}

For every black hole the temperature is given, which would fix the period of the Euclidean time of the soliton. Then, the period of the soliton is fixed in terms of its fluxes through the equation

\begin{equation}\label{period}
\Delta^3 =\frac{\rho_0}{T^3 \omega_0(\nu^2_2=1)}
\end{equation}

$\mathbf{\nu_1^2=1}$. Here
\begin{equation}
\omega_0(\nu^2_1=1)=-\frac{1}{432}\,\left(5+\nu_2^2\right)^{5/2}\left(2+\nu_2^2\right)^{1/2}\left(\nu_2^2+1\right)^{3/2}
\end{equation}

The period of the soliton is again given by the same equation \eqref{period} with the substitution $\omega_0(\nu^1_2=1)\rightarrow \omega_0(\nu^2_1=1)$.

\section*{Acknowledgements} 
This work is supported in part by the FONDECYT grants 1221504, 1230853, 1242043 and 1250133. The work of AA is supported by the Alexander von Humboldt foundation and by the FAPESP grant 2024/16864-9.

\newpage

\hypersetup{linkcolor=blue}
\phantomsection 
\addtocontents{toc}{\protect\addvspace{4.5pt}}
\bibliographystyle{mybibstyle}
\bibliography{bibliografia}

\end{document}